\documentclass[aps,prl,twocolumn,superscriptaddress]{revtex4}

\RequirePackage{lineno}
\usepackage{amsmath,bm}
\usepackage{mathrsfs}
\usepackage{amsfonts}
\usepackage{graphicx}
\usepackage[normalem]{ulem}
\usepackage{epstopdf}
\usepackage{dcolumn}
\usepackage{amsmath}
\usepackage{epsfig}
\usepackage{indentfirst}
\usepackage{psfrag}
\usepackage{subfigure}
\usepackage{amssymb}
\usepackage{color}
\usepackage{setspace}
\usepackage{multirow}

\usepackage[version=3]{mhchem} 
\usepackage{xcolor}
\usepackage{ulem}
\newcommand{\rev}[1]{{\color{black} #1}}
\newcommand{\drev}[1]{{\color{black} #1}}

\begin{document}
	\title{Electrically addressing the spin of a magnetic porphyrin through covalently connected graphene electrodes}

\author{Jingcheng Li}\email{j.li@nanogune.eu}
\affiliation{CIC nanoGUNE, Tolosa Hiribidea 76, 20018 Donostia-San Sebastian, Spain}

\author{Niklas Friedrich}
\affiliation{CIC nanoGUNE, Tolosa Hiribidea 76, 20018 Donostia-San Sebastian, Spain}

\author{Nestor Merino}
\affiliation{CIC nanoGUNE, Tolosa Hiribidea 76, 20018 Donostia-San Sebastian, Spain}
\affiliation{Donostia International Physics Center (DIPC), 20018 Donostia-San Sebasti\'an, Spain}

\author{Dimas G. de Oteyza}
\affiliation{Centro de F{\'{\i}}sica de Materiales 	CFM/MPC (CSIC-UPV/EHU),  20018 Donostia-San Sebasti\'an, Spain}
\affiliation{Donostia International Physics Center (DIPC), 20018 Donostia-San Sebasti\'an, Spain}
\affiliation{Ikerbasque, Basque Foundation for Science, 48013 Bilbao, Spain}

\author{Diego Pe{\~{n}}a}
\affiliation{Centro Singular de Investigaci\'on en Qu\'imica Biol\'oxica e Materiais Moleculares (CiQUS), and Departamento de Qu\'imica Org\'anica, Universidade de Santiago de Compostela, Spain}

\author{David Jacob}\email{david.jacob@ehu.eus}
\affiliation{Departamento de F\'isica de Materiales, Universidad del Pa\'is Vasco UPV/EHU,  20018 Donostia-San Sebasti\'an, Spain}
\affiliation{Ikerbasque, Basque Foundation for Science, 48013 Bilbao, Spain}

\author{Jose Ignacio Pascual}\email{ji.pascual@nanogune.eu}
\affiliation{CIC nanoGUNE, Tolosa Hiribidea 76, 20018 Donostia-San Sebastian, Spain}
\affiliation{Ikerbasque, Basque Foundation for Science, 48013 Bilbao, Spain}

\begin{abstract}

 We report on the fabrication and transport characterization of atomically-precise single molecule devices consisting of a magnetic porphyrin covalently wired by graphene nanoribbon electrodes. The tip of a scanning tunneling microscope was utilized to contact the end of a GNR-porphyrin-GNR hybrid system and create a molecular bridge between tip and sample for transport measurements. Electrons tunneling through the suspended molecular heterostructure excited the spin multiplet of the magnetic porphyrin. The detachment of certain spin-centers from the surface shifted their spin-carrying orbitals away from an on-surface mixed-valence configuration, recovering its original spin state. The existence of spin-polarized resonances in the free-standing systems and their electrical addressability is the fundamental step for utilization of carbon-based materials as functional molecular spintronics systems. \vspace{0.5cm}\\
 
\end{abstract}

 \maketitle

Single molecule spintronics envisions utilizing electronic spins of single molecules  for performing active logical operations. The realization of such fundamental quantum devices relies on the accessibility of electronic currents to the active molecular element, and on the existence of efficient electron-spin interaction enabling  writing and reading information. Single molecule (SM) electrical addressability has been  achieved in break junctions' experiments \cite{Vincent2012,Thiele2014,Gaudenzi2016}, \rev{which realized that the nature of the SM's contacts to source-drain electrodes may affect the ultimate SM functionality } \cite{Lortscher2013}. To ensure stable and reproducible behaviour of the SM device, robust and atomically precise molecule-electrode contacts with optimal electronic transmission are required \cite{Aradhya2013,Xiang2016}. Aromatic carbon systems such as nanotubes, and graphene flakes are considered ideal electrode materials \cite{Ruitenberg2011,Warner2017} because of their structural stability and flexibility, and the large electronic mobility they exhibit. Graphene electrodes also offer the perspective of covalently bonding to a single molecule at specifically designed sites \cite{Auwarter2015}, thus creating robust systems for electrical measurements. 

While  methods to fabricate graphene-SM systems are being developed \cite{Lieaaq0582,Su2018}, the spin addressability by electronic currents through them  remains to be demonstrated. Attractive predictions on the role of graphene-SM hybrids as spin valves, diodes, or rectifiers  \cite{Sanvito2011,Li2016} are  proposed to strongly depend  on the coupling of electronic currents with spin-polarized molecular states and, hence, on the precise contact geometry. Furthermore, spins are sensitive to electronic screening \cite{Oberg2014a} and to electrostatic and magnetic fields \cite{Garcia-Suarez2018} in its local environment. Therefore, precise strategies for spin detection and manipulation in atomically controlled hybrid structures are needed \cite{Li2019} to determine the magnetic functionality of the hybrids and the possible operation mechanisms.

\begin{figure*}[!tb]
\centering
\includegraphics[width = 0.7\textwidth]{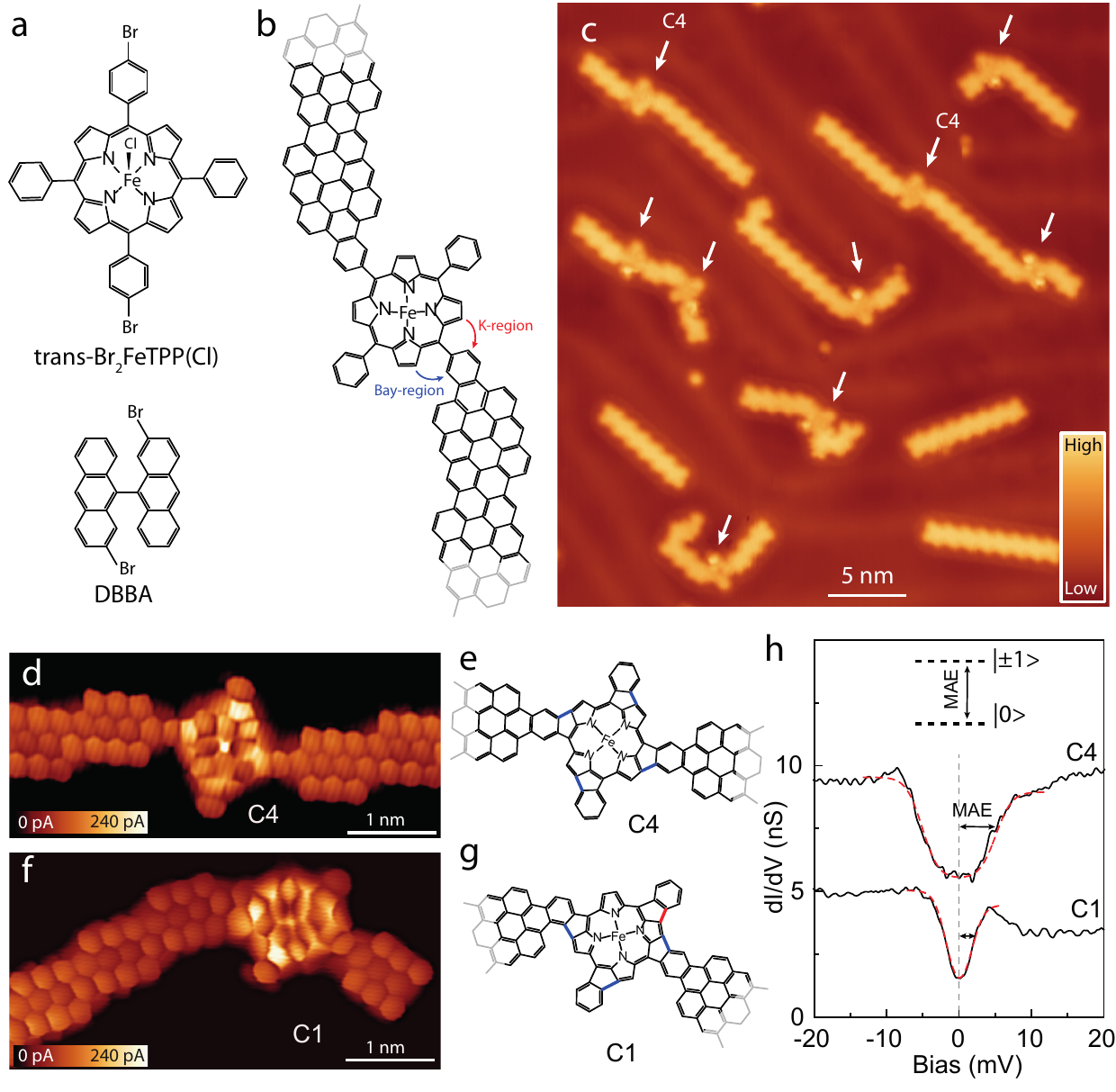}
\caption{\textbf{Synthetic strategy and characterization of magnetic FeTPP contacted by two GNR electrodes in a linear manner.} \textbf{a}, Structures of trans-Br$_2$FeTPP(Cl) (up) and DBBA (down) as the molecular building blocks. \textbf{b}, Structure of a molecular devices after the polymerization and CDH of the two molecular building blocks in \textbf{a}. The CDH can additionally fuse the porphyrin core in a clockwise (K-region, red arrow) or anticlockwise (Bay-region, blue arrow) manner to the contact phenyl. \textbf{c}, STM overview image shows several molecular devices created on a Au(111) surface. The arrows indicate the porphyrins fused to GNR electrodes. \textbf{d,f} Constant height current images measured with a  CO-terminated tip (V=2 mV) of a linear and a bent hybrid structure with a C4- and C1-FeTPP cores, respectively, as shown in  the models \textbf{e,g}, respectively. \textbf{h} Differential conductance spectra over the Fe center of each species, showing inelastic spin excitation steps at the magnetic anisotropy energy (MAE) of the Fe magnetic moment. The inset shows the expected spin excitation for the pristine S=1 FeTPP molecule. Red dashed lines show the fitting to the spectra using the model from Ref.~\cite{Ternes2015}. }   
\end{figure*} 

In this work, we  electrically address the spin of an iron  porphyrin molecule covalently wired to  graphene nanoribbon (GNR) electrodes. To perform  two-terminal transport measurements with  atomic-scale control on the GNR-SM connections, we fabricated a linear GNR-SM-GNR hybrid system utilizing on-surface synthesis (OSS) \cite{Grill2007,Talirz2016,Oteyza:book:2018}. In OSS, pre-designed organic precursors on a flat metal surface are thermally activated to react along specific  polymerization pathways, steering the formation of a targeted molecular extended systems with atomic-scale precision. The obtained linear hybrids were characterized by high-resolution scanning tunneling microscopy (STM) images and compared with results of two-terminal transport measurements of free standing systems bridging the STM tip and sample  electrodes. We demonstrate that the iron spin can be addressed by electronic currents injected into the free standing ribbons. From the transport results, we unveil an intriguing dependence of the spin-excitation spectral fingerprint on the connecting patterns between GNR and SM. For some contact structure the molecular species lie in a mix-valence state on the surface, which vanishes as soon as they are brought into a free-standing configuration. Controlling the alignment of spin-polarized states thus permit operating on the transport mechanisms through such magnetic systems.

\newpage\textbf{Fabrication and study of GNR-porphyrin systems: } 
The linear GNR-porphyrin-GNR hybrid systems were created 
by mixing the two molecular precursors shown in Figure 1\textbf{a} and thermally activating a step-wise reaction on a Au(111) surface. The 2,2’-dibromo-9,9’-bianthracene (DBBA) molecule produces narrow (3,1)chiral GNRs (chGNRs)\cite{Dimas2016}, which are semiconductors with a band-gap of 0.7 eV on Au(111) surface\cite{Merino-Diez2018}. As magnetic species we chose the  Br$_2$FeTPP(Cl) molecule with trans-halogenated configuration to produce linear connections to chGNRs through   surface-mediated Ullmann-like reaction. 
Both molecules were co-deposited onto an atomically clean Au(111) substrate kept at 200$^{\circ}$~C to induce dehalogenation and co-polymerization of the mixture. Subsequently,  the substrate was annealed to 250$^{\circ}$~C to induce a cyclodehydrogenation (CDH) step that forms the planar ribbon structure and, additionally, fuses the tetraphenyl substituents with the porphine pyrroles (Figure 1\textbf{b}) into a fully planar structure\cite{Wiengarten2015,Lieaaq0582} with atomically precise and robust connections.

The overview STM image in Figure 1\textbf{c} confirms that the on-surface reaction proceeds as depicted in Figure 1\textbf{b}, and most FeTPP centers (arrows in Figure 1\textbf{c}) appear connected to two chGNRs   in a trans-configuration. However, not all the fabricated structures appear as straight systems, but in many the terminal chGNRs appear bent with varying angles between them. As we explain in the Figure S1, the different angles appear due to a combination of two structural characteristics: the existence of two different chiral GNR enantiomers on the surface\cite{Merino-Diez2018} and of two bonding configurations of the contacted phenyl with the core pyrroles, either through a K-region or through a Bay-region, as depicted by red and blue arrows in Figure 1\textbf{a}\cite{Wiengarten2015,Lieaaq0582}, respectively.  Straight systems such as the ones labelled C4 in Figure 1\textbf{c}, combine two chGNRs with the same enantiomeric form, both contacting the pyrrole corolla through CDH in the K-region (red arrow in Figure 1\textbf{b}). Bent structures are obtained when different chGNR enantiomers and/or different contact directions (K or Bay) are mixed.

\begin{figure*}[!th]
	\centering
	\includegraphics[width = 0.7\textwidth]{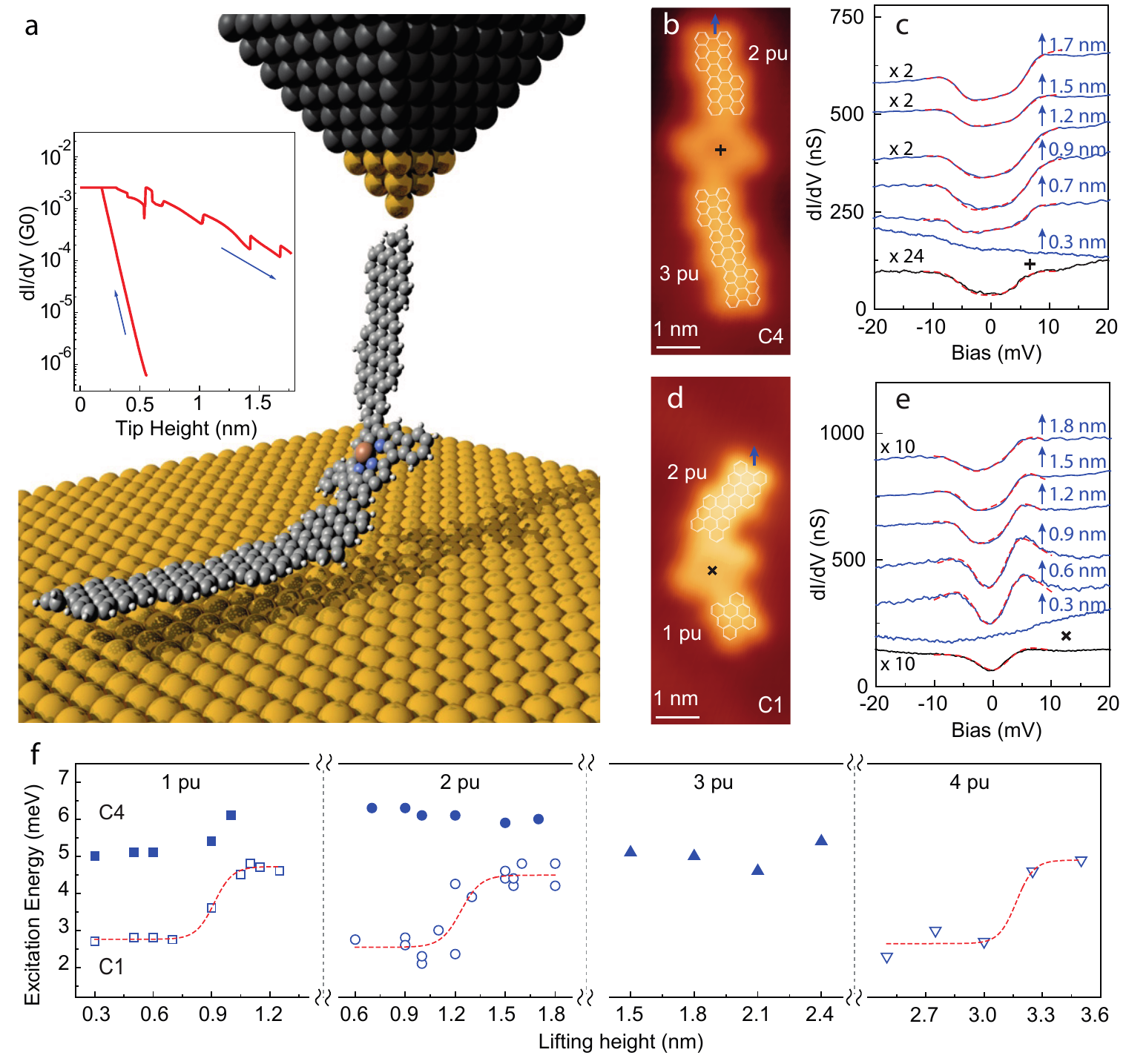}
	\caption{\textbf{ \linespread{1.2}Transport measurements on FeTPP-GNR systems}. \textbf{a}, Illustration of the aimed experimental geometry. Inset shows a typical conductance-distance plot simultaneously recorded during tip 
		approach and retraction. \textbf{b,d}, Constant current STM images of  C4- and C1-FeTPP-GNR systems (V = 0.5 V, precursor units (pu) overlays the figure).
		\textbf{c, e}, dI/dV spectra at different lifting heights (labels) for the two  molecular junctions (spectra vertically shifted for clarity, blue arrows in \textbf{b,d} indicate the contacting points). The  bottom spectra (in black) were measured over the Fe centers, prior to the retraction. Red dashed lines simulate the spectra using the model from Ref.~\cite{Ternes2015}, from which spin excitation energies were obtained. \textbf{f}, Spin excitation energies of the C4 (filled) and C1 (open) FeTPP-GNR systems as a function of lifting heights. The four panels show data from three contacting experiments of each kind, with different GNR lengths (precursor units annotated in the figure). Data points of C1-FeTPP with 1 pu and 2 pu are from several different lifting experiments of the same hybrid system. Red dashed lines are guides to highlight the increase in  excitation energy for C1-FeTPP systems.}   
\end{figure*}

\rev{To determine  the chGNR-FeTPP contact structure} in every case, we measured high-resolution current images using CO functionalized  tips\cite{Gross2009,Hieulle18} (Figure 1\textbf{d,f}).  Most of the straight species appear with all their phenyl rings rotated in the same direction (i.e. C4-symmetric, resulting in characteristic four-fold intra-FeTPP patterns (Figure 1\textbf{e}).  Bent structures frequently show one phenyl ring fused in the opposite direction than the others (i.e. C1-symmetric, Figure 1\textbf{g}).
The precise symmetry of the FeTPP core determines the magnetic properties of the Fe center on the Au(111) surface\cite{Lieaaq0582}.

Differential conductance (dI/dV) spectra measured over the FeTPP core (Figure 1\textbf{h})  show two symmetric steps around zero bias \rev{attributed to inelastic electron tunneling   caused by the excitation of the FeTPP spin 1 multiplet}. As reported previously\cite{Rubio-Verdu2017}, the ground state of FeTPP species on the gold surface maintains the easy-plane magnetic anisotropy (S$_z$=0) of the free molecule, and the dI/dV steps mark the onset of the inelastic excitation into an out-of plane spin configuration (S$_z=\pm$1).\rev{The excitation steps are thus a measurement of the axial magnetic anisotropy constant D of FeTPP molecules.}
On the straight C4 species, the inelastic steps appear at $\pm$ 5.0 mV, corresponding to a magnetic anisotropy energy (MAE) of 5.0  meV. On the bent C1 species, the MAE is smaller (around 1.0 meV), and frequently accompanied by pronounced spectral asymmetries\cite{Lieaaq0582}. 

\textbf{Two-terminal electronic transport measurements: }
We measured the electronic transport through these hybrid GNR-FeTPP-GNR systems by forming junctions as sketched in Figure 2\textbf{a}. These were created by approaching the STM tip to contact the termination of a chGNR  (e.g. blue arrow in Figure 2\textbf{b,d}), and then retracting to lift up the ribbon  from the metal substrate~\cite{Koch2012}. This procedure successfully elevated \rev{the chGNR-FeTPP linear} systems to several nanometers \cite{Chong2016b}. To investigate the effect of electronic currents through the ribbon on the spin states of the FeTPP, we acquired dI/dV spectra during the lifting processes. 

Figure~2 shows the evolution of dI/dV plots with tip retraction height for both straight C4-symmetric (Figure 2\textbf{b,c}) and bent C1-symmetric (Figure 2\textbf{d,e})  GNR-FeTPP-GNR systems. For the C4 symmetric case in Figure 2c, the initial featureless spectra of the contacted GNR termination remain until a lifting height of about 0.7 nm. From this point on, step features emerge in the spectra, resembling the inelastic steps measured over the FeTPP core (black curve in Figure 2\textbf{c}), and remain with this shape during all the retraction expedition.  For comparison, similar dI/dV spectra measured on a bare chGNRs are featureless regardless of the lifting heights (see Figure S2). Hence, we attribute these features  to the inelastic excitation of the FeTPP spin multiplet by inelastic electrons tunneling through the chGNR contacts. We note that  the narrow (3,1) chGNRs are semiconducting\cite{Merino-Diez2018}.  As soon as direct tunneling between tip and sample becomes weak, tunneling through the 0.7 eV chGNR band gap dominates the electrical transport and can reach the FeTPP core, exciting its spin multiplet.

Similar measurements on the bent C1 species also show step-like spectral features for lifting heights larger than  $\sim$ 0.6 nm (Fig,~2\textbf{d,e}). However, in this case  the  step-like features become wider than the reference spectrum (black curve in Figs. 2\textbf{e}), and develop an asymmetric component, \rev{a characteristic conductance rise at the positive step, for several retraction values. These are characteristic of finite amount of potential scattering \cite{Rubio-Verdu2017}, and can be simulated  using a high-order scattering model \cite{Ternes2015} or, as we shall see later, under the framework of the Anderson  model. }
Figure~2\textbf{f}  summarizes the  MAE values extracted from spectra  at different lifting heights \rev{(see Supplementary Information)}, for both C1-FeTPP and C4-FeTPP connected to lifted GNRs of varying lengths (quantified by the number of precursor units (pu)). Focusing on the system presented in Figure 2d with 2 pu, the MAE (empty symbols) 
shows a step-wise increase from the original $\sim$2.5~meV to $\sim$4~meV at a retraction  height of $\sim$1.2 nm, and remains constant around this value henceforth. 
A similar   step-wise  increase of the MAE was found in lifting experiments of other C1-FeTPP systems  with  different ribbon's lengths (e.g. 1 pu, and 4 pu in Figure 2\textbf{f}). They appear at lift heights scaling with the length of the ribbons (e.g. 3 nm for ribbons consisting of four precursor units), indicating that they are caused by the raise of the  FeTPP core from the metal surface, to bridge the tip and sample (Figure 2\textbf{g}).
In contrast, the  MAE of C4-FeTPP systems remains constant throughout the whole lifting process, thus not being affected by the FeTPP detachment.  These results indicate a hidden mechanism  controlling the MAE for the case of the asymmetric C1-FeTPP cores.

\begin{figure}[!tb]
\centering
\includegraphics[width = 0.48\textwidth]{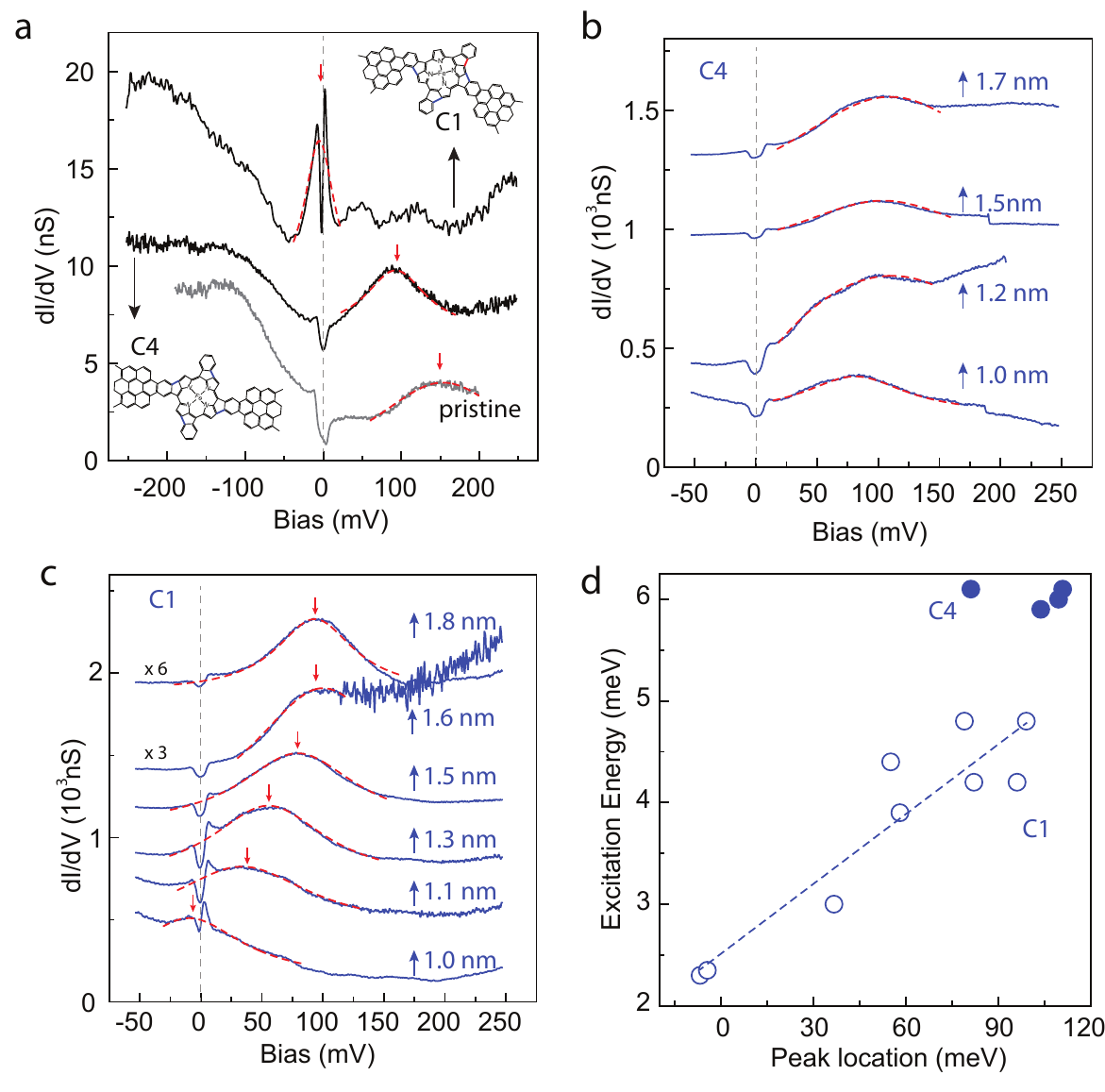}
\caption{\textbf{Effect of d orbital filling on the MAE of FeTPP}.  
\textbf{a}, dI/dV spectra recorded on a pristine FeTPP (saddle shaped, contacted to  chGNRs), planar C4- and C1-FeTPP species on the bare surface.  
\textbf{b,c}, dI/dV spectra recorded at different lifting heights of C4- and C1-FeTPP respectively. The lifted hybrid systems are the same as the ones in Figure 2\textbf{c,e}.  The spectra in \textbf{a-c} are shifted for clarity, and the red dashed lines show the Lorentz fitting to the resonances of different types of FeTPP. Red arrows indicate the center of the peaks. \textbf{d}, Excitation energy (i.e. MAE) extracted from the inelastic features using the model from Ref.~\cite{Ternes2015} as the function of energy location of resonances  in \textbf{b,c}.}  
\end{figure}
 
To shed light on the origin of the different MAE evolution with elevation of the C4- and C1-FeTPP species, we investigated the effect of lifting on the spin carrying orbitals by recording the dI/dV spectra over wider bias range. The magnetic moments of transition metal porphyrins are distributed in a set of $d$ orbitals, split according to the ligand field around the Fe ion. For pristine FeTPP, two spin-polarized frontier orbitals around E$_F$ account for the S=1 magnetic ground state on a Au(111) surface \cite{Karan2018,Rubio-Verdu2017,Jacob_2018}.   
Figure 3\textbf{a} compares dI/dV spectra of pristine FeTPP (chGNR contacted) molecules with that of C4- and C1-FeTPP hybrids on the bare surface. The spectrum on  C4-FeTPP reproduces the frontier states of pristine FeTPP but shifted to lower energy: \rev{a state at negative bias, associated to part of the d-manifold of the Fe ion,  and the state at positive bias, attributed to the singly unoccupied (SU) state\cite{Rubio-Verdu2017}. Interestingly, on}   the C1-FeTPP system, the SU-resonance is shifted further to now overlap with E$_F$. Wide-range  spectra measured while lifting the FeTPP-GNR hybrids show that the SU-state of the C4-FeTTP remains around similar values during the whole tip retraction process (Figure 3\textbf{b}). In contrast, the resonance centred at E$_F$ in the C1-FeTPP species  shifts   away from the Fermi level with the lifting height, until a value of around 100~mV is reached (Figure 3\textbf{c}). Interestingly, the MAE values extracted in Figure 2\textbf{f} show a linear dependence with the energy values of SU-peaks (Figure 3\textbf{d}), suggesting that both effects are related.

\begin{figure*}[!t]
\centering
\begin{tabular}{cc}
 \includegraphics[width = 0.4\textwidth]{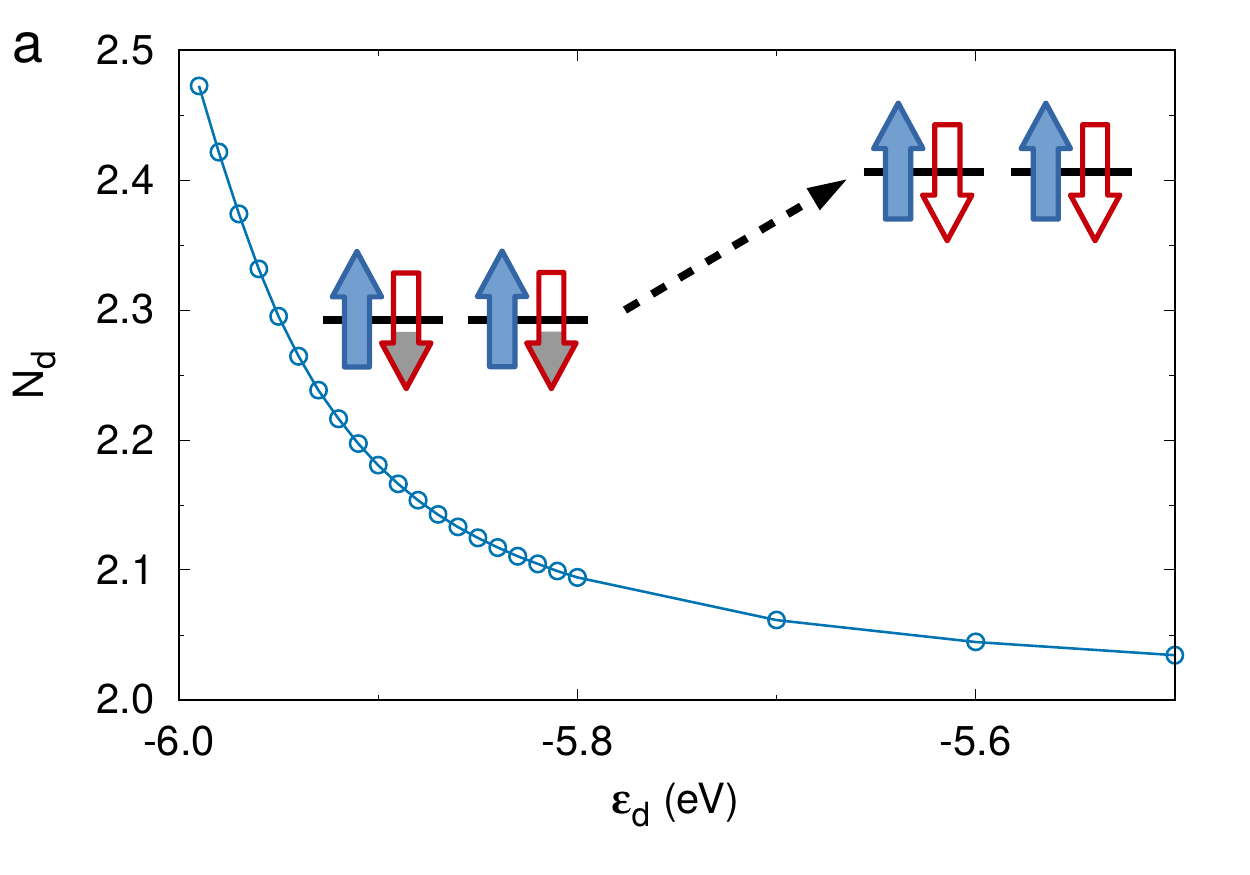} &
 \includegraphics[width = 0.4\textwidth]{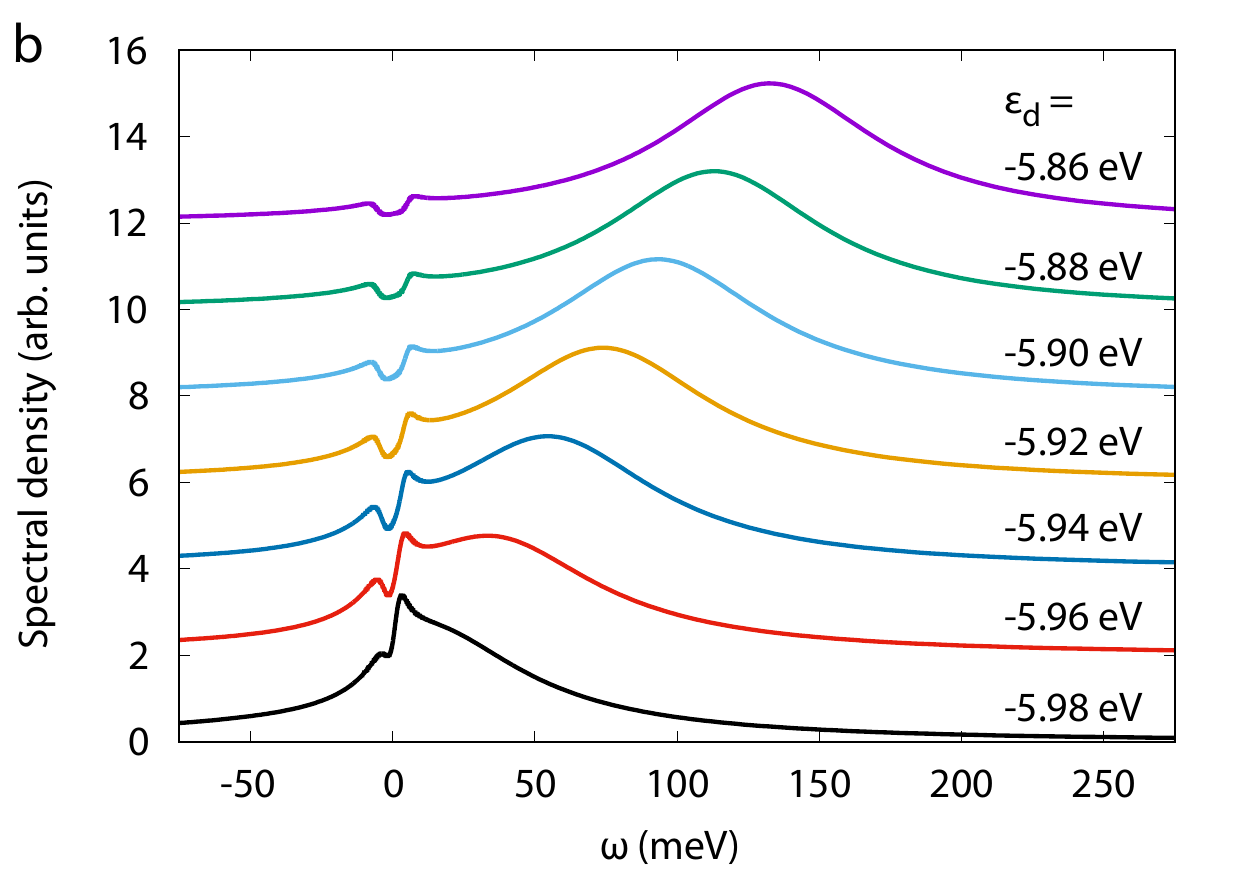} \\
 \includegraphics[width = 0.4\textwidth]{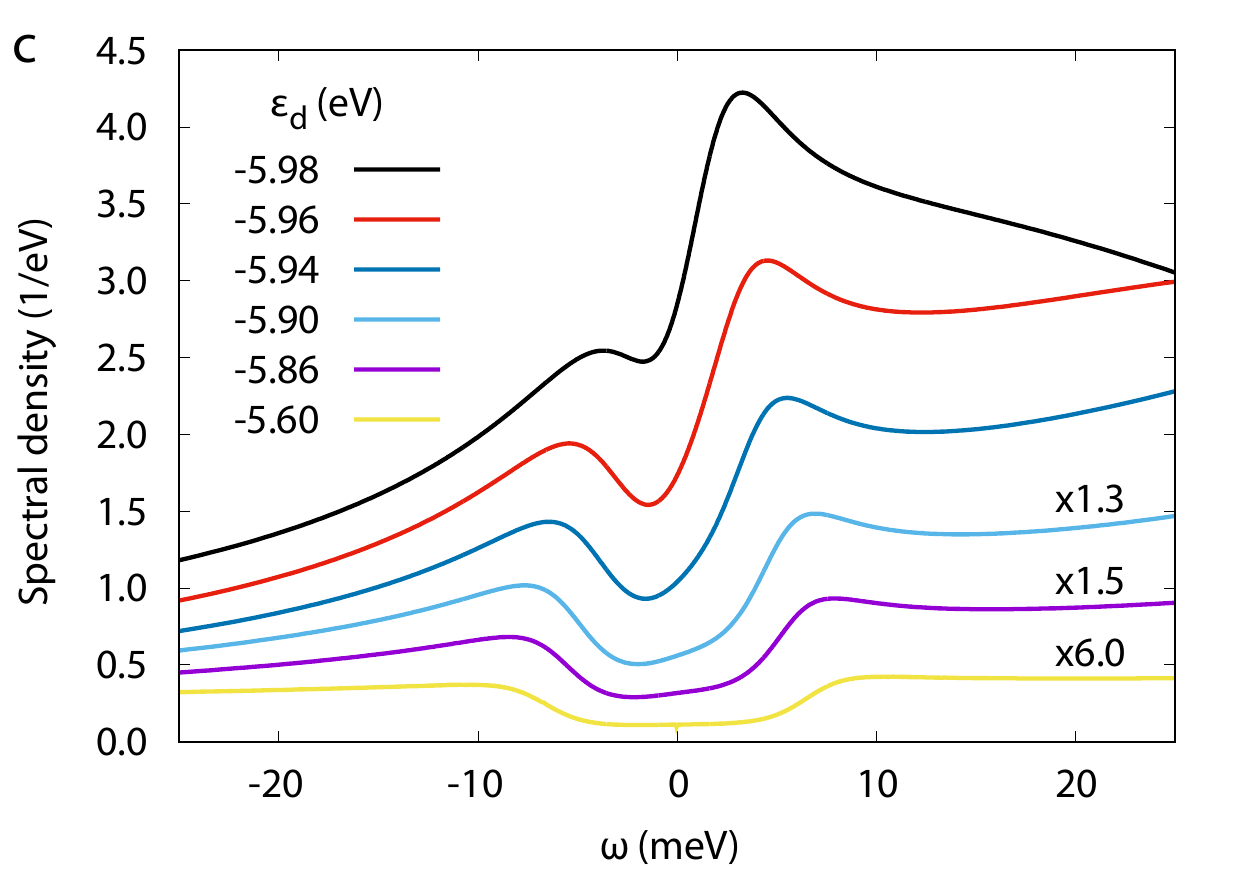} &
 \includegraphics[width = 0.4\textwidth]{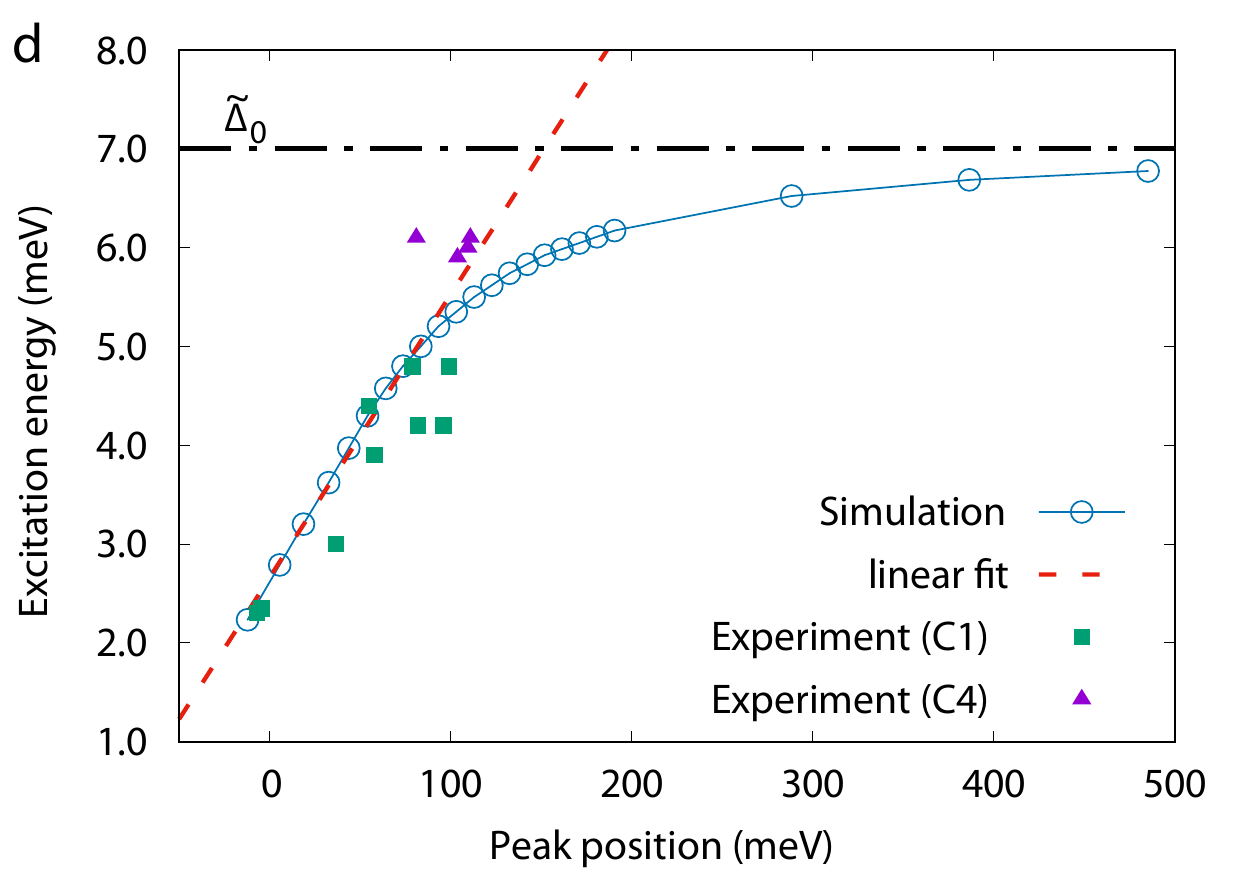} 
 \end{tabular}
\caption{\textbf{Two-orbital Anderson model simulations}. The simulations were done with uni-axial magnetic anisotropy $D=7.14$meV, single-particle broadening $\Gamma=30$meV, intra- and inter-orbital Coulomb repulsion $U=3.5$ eV and $U^\prime=2.5$ eV, respectively, Hund's rule coupling $J_H=0.5$ eV
\drev{and temperature $T\sim10 K$}.
\textbf{a}, Impurity occupancy $N_d$ as a function of the on-site energy $\varepsilon_d$.
\textbf{b,c}, Spectral functions for different on-site impurity energies $\varepsilon_d$   show  the upper Coulomb peak shifting (\textbf{b}) and in a smaller energy
window around Fermi level,  the resulting inelastic step features (\textbf{c}).
\textbf{d} Excitation energy as a function of the Coulomb peak position. The experimental data from Figure 3\textbf{d} are also included here for comparison.
}  
\end{figure*}

\textbf{Theoretical modelling of MAE renormalization: } 
The shift of a molecular state away from the Fermi level as the ribbon is lifted \sout{up} indicates a change in the filling of the spin-carrying $d$-orbitals of the C1-FeTPP core. Thus, a plausible explanation for the observed dependence of excitation energy on resonance alignment is the renormalization of the MAE by charge fluctuations \drev{recently predicted by one of us} \cite{Jacob2018}.
In order to verify this scenario we modeled the spin excitations of the S=1 system coupled to the surface by a two-orbital Anderson impurity model (2AIM) close to half-filling including a uni-axial magnetic anisotropy term \drev{$D S_z^2$, that we solve} in the one-crossing approximation (OCA)\cite{Haule2010}. 
\drev{More details about the model and its solution can be found in the Supplementary Information and in previous works \cite{Jacob2016,Jacob2018}. 

Close to half-filling ($N_d\sim2$) the ground state (GS) of the impurity is the $S=1$ triplet state. As we mentioned above, the large positive magnetic anisotropy ($D>0$) of FeTPP splits the $S=1$ multiplet into a non-degenerate $S_z=0$ GS and a $S_{\pm}$ excited state, spaced by the bare MAE $\Delta_0=D$. 
The low energy part of the spectrum ($|\omega|<10$meV) shows the typical steps close to $\pm{D}$ characteristic of inelastic spin-flip excitations of a spin-1 quantum magnet broadened by hybridization with the substrate (yellow curve in Figure 4\textbf{c}). }

\drev{In this scenario, spin-fluctuations caused by the Kondo effect are not possible due to the lack of GS degeneracy. However, charge fluctuations are still possible and become important as electron occupation changes and the system approaches towards a mix-valence regime, where MAE is renormalized and can acquire asymmetric line-shapes \cite{Jacob2018}. }
We simulated changes in orbital filling  by shifting the impurity levels $\varepsilon_d$ downwards, resulting in the increase of electron occupation $N_d$ of the $d$-orbitals (Figure 4\textbf{a}). 
\drev{Correspondingly,} the model spectra (Figure 4\textbf{b}) show a pronounced peak at positive energies that moves towards the Fermi level as the impurity levels are lowered in energy and their filling increases. This peak is the upper Coulomb peak of the Anderson model, \drev{and corresponds}  to the SU state of our \drev{experimental} spectra \drev{(cf. Figure 3\textbf{b,c})}.
For $\varepsilon_d\sim-6$eV the resonance moves directly over E$_F$ and the system enters the mixed-valence regime ($N_d\sim2.5$). This state is characterized by strong charge fluctuations that lead to a correspondingly strong renormalization of the bare MAE ($\Delta_0$).
The resulting spin excitation features around zero (black line in Figure 4\textbf{c}) are fainter and narrower, and with asymmetric components, similar to the experimental plots on C1-FeTTP species on the surface (cf. Figure 2\textbf{e}).

\drev{Our 2AIM results suggest that the } lift process \drev{effectively} gates the SU resonance of the C1-FeTPP core away from E$_F$, driving the system towards half-filling ($N_d\sim2$), thus reducing the charge fluctuations. This leads to an increase in the MAE, approaching towards $\Delta_0$, and to more symmetric inelastic steps. 
The variation of spectral features with the alignment of the impurity level shown in Figure 4\textbf{c} reproduces the evolution of spin excitation spectra of C1-FeTPP species with the lift height observed in the experiments (cf. Figure 2\textbf{e}). 

The 2AIM also reproduces the correlation between MAE and orbital alignment found in the experiments (Figure 4\textbf{d}). The MAE shows a linear behavior for peak positions up to 100 meV, while for larger values  the MAE develops a sublinear behavior, approaching the upper limit given by  $\tilde{\Delta}_0$ at the particle-hole symmetric point ($\varepsilon_d=-4$eV). At this point, charge fluctuations are fully suppressed ($N_d=2$). Still, the value $\tilde{\Delta}_0$ is slightly smaller than the bare MAE of $\Delta_0=D=7.14$meV due to renormalization by Kondo exchange coupling with the conduction electrons\cite{Oberg2014a,Jacob2018}. 

\textbf{Discussion: } 
The excellent agreement between calculations and  experimental results suggests that changes in the filling of the $d$-orbitals are \rev{the main} responsible for the variations of MAE in planar FeTPP cores embedded in chGNRs\cite{Lieaaq0582}. \rev{The asymmetric species C1-FeTPP lie in a mixed-valence state on the surface, where valence fluctuations strongly distort spin signal and renormalizes the effective spin anisotropy. Detaching the C1-FeTPP core from the surface removes the charge transfer and brings the system closer to a half-filling state, with stepped spin-excitation signal and larger anisotropy.}
In contrast, the fairly constant MAE value of the C4-FeTPP species at all lift heights \rev{agrees with} the small variation of their $d$-orbitals during retraction. \drev{As shown in Figure 3\textbf{a}, the SU peak of these species lies well above  the Fermi level ($\sim90$ meV) already before lifting, hence indicating that the corresponding $d$-orbitals are close to half-filling already on the surface. The lack of MAE variations of C4-FeTPP molecules during the lift process also proves that molecular structural distortions are minor during lifting, since these would cause significant MAE variations \cite{Heinrich2015}. } 

\rev{The different occupation state of each species on the surface is probably associated to their different core structure. We note that hybridization with the substrate, which can control orbital filling, is very sensitive to small changes of the ligand environment of the Fe ion. In particular, the lower symmetry of C1 species probably forces the Fe ion out of the molecular plane and, hence, make it more prone to interact with the metal substrate. We cannot neglect that different configuration of orbital filling caused by the different core structure are also affecting the degree of orbital overlap with the substrate and, hence, the charge transfer.  }

\rev{A further intriguing aspect is that the MAE values reached by lifted C1-FeTPP systems are still  smaller than in C4-FeTPP species, even when both have similar orbital alignment (c.f. Figure 3\textbf{d}).}  The lower axial symmetry of the asymmetrically fused species, which causes a larger degree of orbital mixing, is a probable additional mechanism accounting for a lower intrinsic MAE of C1-FeTPP systems\cite{Parks2010,Bryant2013b}.

The survival of the molecular spin in the free-standing graphene systems, together with  the presence of spin-polarized orbitals around the Fermi level, are key ingredients for  realizing an  organic  spin-filter device\cite{Cho2011}. We envision that, combined with spin-polarized currents (e.g. from magnetic metal electrodes) and magnetic fields, many of the predicted functionalities of these model platforms as molecular spintronics elements  could be experimentally tested, benefiting from the high degree of reliability of its covalent construction.  

\textbf{Methods}\\
The experiments were performed in a custom-made low-temperature STM, at low temperatures (5 K), and under ultrahigh vacuum conditions.  The molecular building blocks 2,2'-dibromo-9,9'-bianthracene(DBBA, prepared as described by de Oteyza et al. \cite{Dimas2016}) and trans-Br$_2$FeTPP(Cl) (from PorphyChem SAS) were simultaneously sublimated from separate Knudsen cells onto  a clean  Au(111) surface (cleaned by cycles of Ne+ ion sputtering and annealing at 730 K) at room temperature by
thermal sublimation from quartz crucibles (sublimation temperatures,
TDBBA = 170$^\circ$C and trans-Br$_2$FeTPP(Cl)= 310$^\circ$C). The ratio between DBBA and trans-Br$_2$FeTPP(Cl) was kept at around 4 to 1. Subsequently, the sample was further annealed at 250 $^{\circ}$C for 5 minutes to induce the on-surface synthesis of the hybrid nanostructures.    A tungsten tip was used in the experiment. High resolution constant-height current images were acquired with a CO-functionalized tip at very small voltages, and junction resistances of typically 20 M$\Omega$.
The CO molecule was removed for tip-GNR junction formation.
The $dI/dV$ signal was recorded using a lock-in amplifier with a bias modulation of $V_\mathrm{rms}=0.4$ mV (spectra in Figure 2) and 1 mV (spectra in Figure 3) at 760 Hz respectively. Analysis of STMdata was performed with the WSxM software \cite{Horcas2007}. Retraction heights are labelled respect to a tip-substrate contact point, obtained for every tip from tip-surface approach test experiments.

\textbf{Acknowledgements}: We thank Manuel Vilas-Varela for the synthesis of the chGNR molecular precursor. We are indebted to Carmen Rubio for fruitful discussions. We acknowledge financial support from Spanish AEI (MAT2016-78293-C6 and the Maria de Maeztu Units of Excellence Programme MDM-2016-0618),  the EU project PAMS (610446), the Xunta de Galicia (Centro singular de investigaci\'on de Galicia  accreditation 2016-2019, ED431G/09),   and the European Regional Development Fund (ERDF).  DJ acknowledges funding by the Ikerbasque foundation through a starting grant.

\bibliography{reference}

\end{document}